\documentclass[12pt]{article}
\usepackage{a41}
\usepackage{epsfig}

\begin{document}

\begin{center}
{\LARGE\bf  A study on $\Delta u$ and $\Delta d$ 
in charged current events
using polarized beams at HERA}

\vspace{1cm}
{J.\ G.\ Contreras$^a$,  A.\ De\ Roeck$^b$, M.\ Maul$^c$}

\vspace*{1cm}
{\it $^a$Universit\"at Dortmund, Institut f\"ur Physik, D--44221
Dortmund, Germany}\\

\vspace*{3mm}
{\it $^b$Deutsches Elektronen--Synchrotron DESY, Notkestra\ss e 85,
D--22603 Hamburg, Germany}\\

\vspace*{3mm} 
{\it $^c$Institute for Theoretical Physics, Universit\"at
Regensburg, 93053 Regensburg, Germany}\\ 

\vspace*{2cm}

\end{center}

\begin{abstract}
Charged current events are studied at HERA with polarized
beams, using a Monte Carlo
event generator and  taking into account detector smearing effects.
The observed asymmetries are used to study the 
 extraction of the polarized structure function
$g_5$ and the polarized parton distributions
$\Delta u$ and $\Delta d$ in the proton.
\end{abstract}

\section{Introduction\label{sec_intro}}

\vspace{1mm}

The flavour decomposition of the polarized quark distributions in 
the proton is  important information for our understanding of the spin
structure of the proton. At present fixed target experiments 
the 
asymmetries measured in semi-inclusive data~\cite{smc}
 have been exploited for
this purpose, a method which can also be used at 
 HERA with polarized beams~\cite{maul}. This option for  HERA
 has been proposed in~\cite{workshop}, and leads to collisions of
 27.5 GeV polarized electrons on 820 GeV polarized protons.
The high center of mass energy of HERA ($\sqrt{s} = 300 $ GeV) allows
in addition to study  charged current (CC) events which probe 
different combinations of spin-dependent quark distribution functions 
compared to  
neutral current interactions. In particular, CC interactions
 distinguish quark from 
anti-quark flavours.

This paper is a continuation of the analytical 
calculations  reported in \cite{ansel}, and is based on a 
Monte Carlo study of CC events. For this purpose the  Monte Carlo
event generator PEPSI 6.5~\cite{PEPSI1,PEPSI65} was  
upgraded to contain the full electroweak 
structure at tree level for lepton-proton scattering. It was used
to generate event samples which were tracked through a detector 
program to study detector smearing effects. PEPSI is a 
version of the program LEPTO~\cite{lepto}, extended for polarized 
beams and targets. It contains leading order
QCD matrix elements, supplemented with parton showers to approximate
higher order effects. 
The hadronization in PEPSI is based on the Lund string fragmentation
model~\cite{lund}. The cross section results of PEPSI have been checked
with the analytical calculations reported 
in~\cite{ansel}, and were found to be in 
agreement.

The sensitivity of the CC cross section to the polarized quark 
distributions is shown in the following.
The hadronic scattering tensor for lepton-nucleon interactions 
can be decomposed in three unpolarized and five polarized structure 
functions. The latter ($g_1$ to $g_5$) are not independent. Assuming
$g_4 \sim 0$, which is strictly true in the parton model, the differential
cross section for CC reactions on a  longitudinally polarized target 
reads~\cite{ansel}:
\begin{equation}
\frac{d^2\sigma}{dxdQ^2}  = 
\frac{G^2_FM^4_W}{4\pi}\frac{1}{(Q^2+M^2_W)^2}
 ([aF_1(x,Q^2)-\lambda b F_3(x,Q^2)] + [-2\lambda b g_1(x,Q^2) + 
ag_5(x,Q^2)]\delta),
\label{eq_cs}
\end{equation}
with $x$ being Bjorken-$x$, and $Q^2$ the four-momentum transfer squared
in the deep inelastic scattering process.
$G_F$ is the Fermi constant, and $M_W$ is the $W$-Boson mass; 
$\lambda$ is $-1 (+1)$ for the $e^-$ 
($e^+$); $\delta$ denotes the longitudinal polarization of the nucleon 
to be anti-parallel (+1) or parallel ($-1$)
to the polarization of the lepton;
$a = 2(y^2-2y+2)$
and $b = y(2-y)$, with $y$ the usual inelasticity variable.
The asymmetry  is defined by
\begin{equation}
A^{W\mp} =
\frac{d\sigma^{W^\mp}_{\uparrow\downarrow}-d\sigma^{W^\mp}_{\uparrow\uparrow}}
{d\sigma^{W^\mp}_{\uparrow\downarrow}+d\sigma^{W^\mp}_{\uparrow\uparrow}}
\nonumber\\
 = \frac{\pm 2bg^{W^\mp}_1+ag^{W^\mp}_5}{aF^{W^\mp}_1\pm
   bF^{W^\mp}_3}, \label{eq_as}
\end{equation}
where in leading order
\begin{equation}
\begin{array}{ccc}
g^{W^-}_1 = \Delta u+\Delta c + \Delta\bar{d}+ \Delta\bar{s}, 
& \mbox{   }&
g^{W^+}_1 = \Delta d+\Delta s + \Delta\bar{u}+ \Delta\bar{c},\\
g^{W^-}_5 = \Delta u+\Delta c - \Delta\bar{d} - \Delta\bar{s}, 
& \mbox{   }&
g^{W^+}_5 = \Delta d+\Delta s - \Delta\bar{u} - \Delta\bar{c},\\
F^{W^-}_1 =  u+ c + \bar{d}+ \bar{s}, 
&\mbox{   }&
F^{W^+}_1 =  d+ s + \bar{u}+ \bar{c},\\
F^{W^-}_3 = 2( u+ c - \bar{d} - \bar{s}), 
&\mbox{   }&
F^{W^+}_3 = 2( d+ s - \bar{u} - \bar{c}).\\
\end{array}
\end{equation}
Here $\Delta q $ denotes  the polarized parton distribution
which is the difference between the probability to find a parton in the
 longitudinally polarized 
proton whose spin is aligned minus the one whose spin is anti-aligned,
and $q$ is  the unpolarized parton
distribution.

To evaluate the measurability and quality of the spin asymmetries for
 CC events
several systematic studies have been  performed. 
A total luminosity of 500 pb$^{-1}$ and a total polarization
of 0.5 (0.7 for each beam)  was assumed and generated
with PEPSI, compatible with the expectations for HERA~\cite{workshop}.
 The polarized parton distributions GS--A~\cite{gs} and 
the unpolarized parton distributions GRV~\cite{grv} have been used
throughout this analysis.
The available
luminosity was distributed in equal parts for  incoming  electron
and positron beams, and for parallel or antiparallel
polarizations; i.e., for each case 125 pb$^{-1}$ was used.
However, since 
 the proton has roughly two times more up than down quarks,
it could be better to assign two thirds of the luminosity to the $W^+$
case to obtain results with equal statistical
significance. Since this does  not interfere with other 
polarized measurements such a
scenario should be  considered for future measurements and studies.

\section{Detector Effects and Simulation}

\vspace{1mm}
\noindent  
To investigate the measurability of the spin asymmetries a fast
simulation of a HERA type of detector~\cite{h1,zeus} has been
implemented \cite{detector} (for more details see for example
\cite{scheins}). The experimental characteristic of charged current
interactions at HERA is a high $p_t$ neutrino in the final state
leading to large missing transverse momentum for the detected event.
The corresponding selection of the events relies essentially on the
calorimetric measurement of the hadronic final state.  For this study
we based the simulation on the calorimeter of the H1 experiment. The
simulation parameters have been tuned to real data, and some details
will be given in the following.

A calorimeter with a separate backward and  central/forward part was
considered.  Backward here means the electron 
direction. The calorimeter covers the polar angle $\theta$ (defined
with respect to direction of the incoming proton) from 4$^\circ$ to
177$^\circ$. The central/forward and backward parts are separated at 
152$^\circ$. An optional acceptance in the range
$10^\circ<\theta<170^\circ$ was implemented to take into account the
planned luminosity  upgrade of  HERA~\cite{upgrade}, which would reduce the 
acceptance as a result of machine magnets which are installed close to the 
interaction point.
The complete range in  azimuthal angle $\phi$ is
covered.

The influence of the inactive material in front of the calorimeter, and 
of losses due to the calorimeter  edges and in the
region between the central/forward and backward part was modeled with a set of
efficiency factors which depend on the polar angle of the simulated
particle. 

The smearing of the energy and 
reconstructed angle of the detected particles was
applied as follows. The energy resolution of
the backward part is 
taken to be $\sigma_{back}/E=0.56$, and of the central part
$\sigma_{cent}/E=0.46/\sqrt{E}\oplus0.73/E\oplus0.026$, where the
energy is given in GeV.  For the angular resolutions 
$\sigma_\theta= 50$ and $\sigma_\phi = 90$ (in mrad) was used.

Further, a Gaussian smearing (with RMS 11 cm) of the interaction 
vertex position along the beam line, 
consistent with current HERA operation, was
included in the simulation.

The efficiency to trigger and select charged current events was set to
85$\pm$3\% in accordance with current estimates~\cite{h1cc}
for charged current events. The
background of photoproduction events, cosmic rays and production of
$W^\pm$ or Z$^0$ Bosons is very small in the present  analyses 
at HERA  and
was ignored in the following~\cite{h1cc, zeuscc}.

We will study the asymmetries as a function of Bjorken-$x$.
The bin size in $x$ as used in the following  was chosen taking 
into account statistics and 
event migration due to detector effects. For each
bin we have a 
purity\footnote{number of events generated and simulated in a bin
  divided by the number of simulated events in the same bin} and an
efficiency\footnote{number of events generated and simulated in a bin
  divided by the number of generated events in the same bin} bigger
than 80\%. 

\section{Results}
   
\subsection{Kinematics}

\vspace{1mm}
\noindent

The H1 collaboration selects CC events requiring a minimal missing
transverse momentum, $P_{Tmiss}$, of the hadronic final state to be larger
than 25 GeV and presents results with a minimal $Q^2$ of 625 GeV$^2$
\cite{h1cc}. The ZEUS collaboration on the other hand has recently
published an analysis of CC events with the selection criteria of
$P_{Tmiss}>11$ GeV and $Q^2>200$ GeV$^2$ \cite{zeuscc}. The lower cut
allows to extent the reach in kinematics from 
$\log_{10}(x)\approx-1.5$ to
$\log_{10}(x)\approx-2.0$$\;.$

\begin{figure}[!hp]
  \centering \epsfig{file=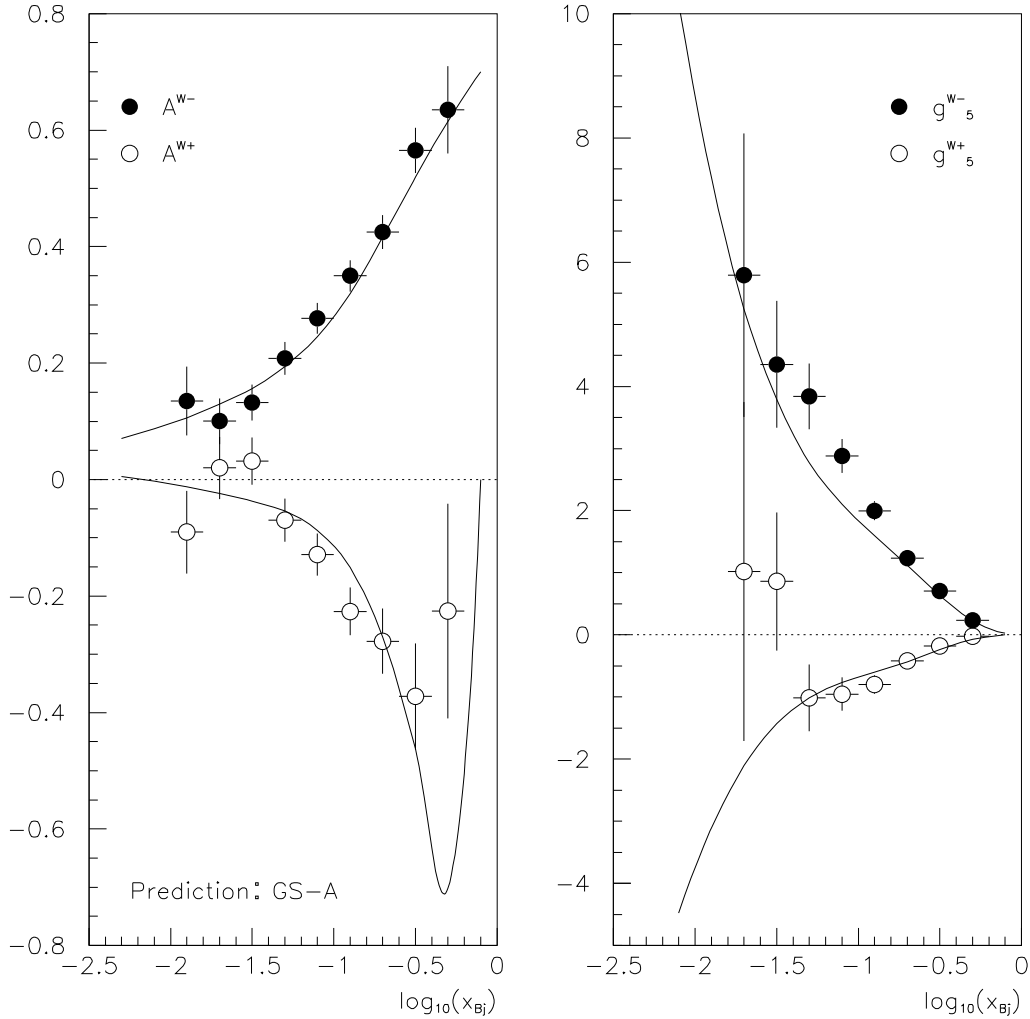,width=.9\textwidth}
\caption{Spin asymmetries $A^{W^-}$ (full symbols, left side) and $A^{W^+}$
  (open symbols, left side) for CC events are presented for a total
  luminosity of 500 pb$^{-1}$. Also shown are the structure functions
  $g^{W^\pm}_5$ (right side) extracted from the asymmetries. The parton
  densities GS--A from \cite{gs} were used. The error bars represent
  the statistical uncertainty of the measurement.} 
\label{fig_results}
\end{figure}

For the studies presented in this contribution $P_{Tmiss}>15$ GeV and 
$Q^2>225$ GeV$^2$ have been used, which seem a reasonable
assumption based on the present day experience at HERA.
The results for the asymmetries, including detector effects,
 are shown on the left side of figure
\ref{fig_results}. The error bars indicate the statistical precision 
of the measurement. The asymmetries are very large, as noticed before 
in~\cite{ansel}, so that the data allow for a significant 
measurement.
The solid line is the result of the analytical calculation of the 
asymmetry using the formulae of section 1, and the same parton 
distributions as used for the 
 PEPSI generated events. It shows that the detector smeared 
asymmetries  
are in good agreement with the true ones.

Note that in our kinematic domain we have $a>>b$,
thus the asymmetry to a good approximation directly measures the 
structure function ratio $g_5/F_1$:
\begin{equation}
A^{W^{\pm}}\approx \frac{g^{W^{\pm}}_5}{F^{W^{\pm}}_1}
\label{eq_as_app}
\end{equation}
The structure functions $F^{W^{\pm}}_1$ will be already precisely
measured by the time the  analysis of polarized data 
from HERA can be performed.
This makes it possible to extract the structure functions $g^{W^{\pm}}_5$
from the measured asymmetries. Here we multiply the asymmetry with 
$F^{W^{\pm}}_1$ as calculated from the GRV distributions.
The results are presented on the right
side of Fig.~\ref{fig_results} and 
 compared with  the asymmetries from  the analytical 
calculation for $g_5$. 
It shows that this approximation works well (to the 10-20\% level)
in our kinematic range.

\subsection{Energy scale of the calorimeter}

\vspace{1mm}
\noindent
The selection and measurement 
of CC events depends strongly on the calorimeters of
the detector. This implies that one of the dominant sources of
systematic uncertainty of the measurement is the precision on the
knowledge of the energy scale of the calorimeters. 
The electromagnetic energy scale will be rather well determined for the 
complete detector, to better than 1\%, but the uncertainty on the 
hadronic energy scale will be larger.

\begin{figure}[!hp]
\centering
\epsfig{file=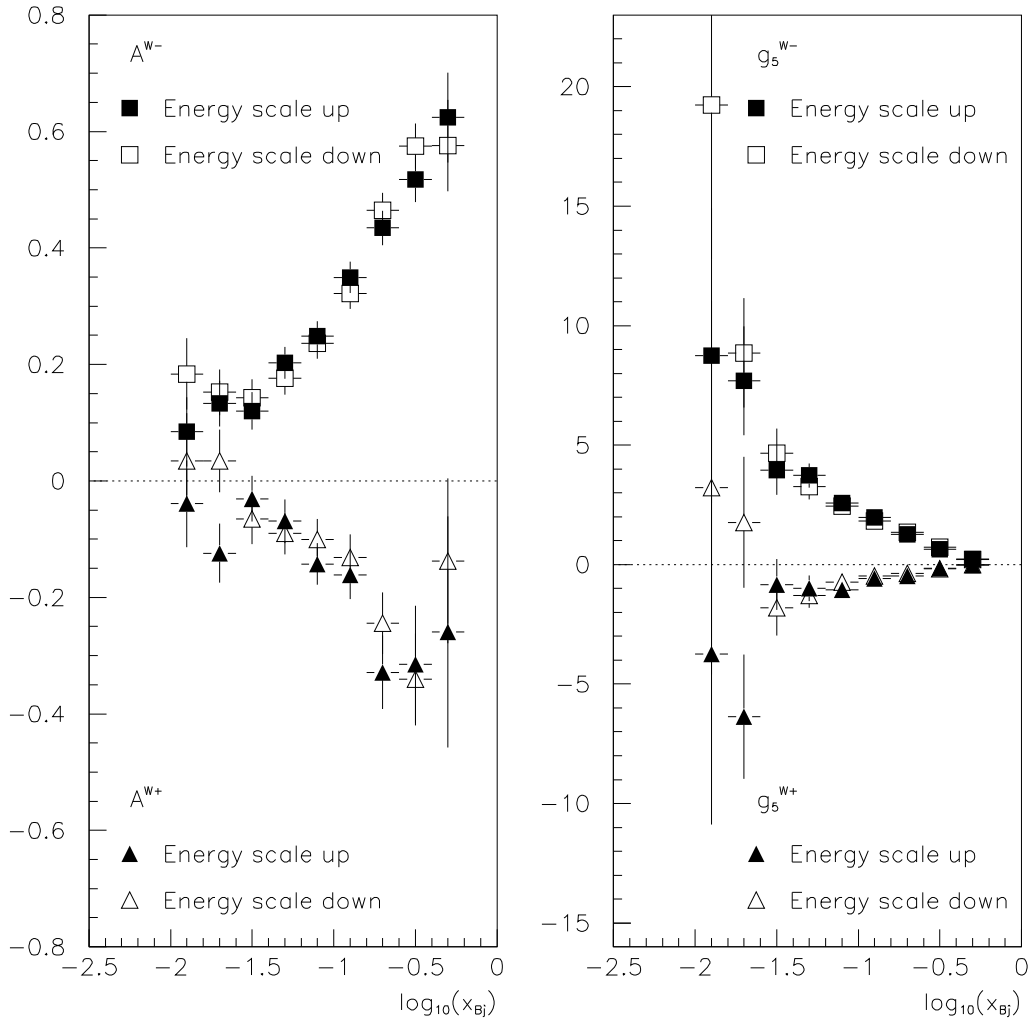,width=.9\textwidth}
\caption{Effect on $A^{W^\pm}$ (right) and $g^{W^\pm}_5$ (left) of the
  variation of the absolute energy scale of the central/forward and backward
  calorimeters.}
\label{fig_en_sc}
\end{figure}

To study this effect the energy scale of the calorimeters was changed
in the simulation program by $\pm$ 10\% in the backward part and by $\pm$
4\% in the central/forward region. These values are in agreement with those
currently used by the H1 experiment and thus represent a somewhat
pessimistic scenario. We find  that the
effect of varying the energy scales on  the
asymmetries and the structure function $g_5$ is within the
statistical fluctuations of the measurement, shown in Fig.~\ref{fig_en_sc}. 
Hence, it turns out that  the effect of the
energy scale uncertainty largely cancels in the measured ratio of 
cross sections (see eqn.~2).

\subsection{Acceptance of the detector}

\vspace{1mm}
\noindent
Currently the geometrical acceptance of the 
central detectors at the HERA collider is
 about 4$^\circ$ $-$ 177$^\circ$ in  polar angle
$\theta$. This will be reduced by the high luminosity upgrade
of HERA.
For this upgrade it is planned  to install
magnets within the detectors, close to the 
interaction point, tightly surrounding the beam pipe~\cite{upgrade}. 
Apart from a reduced angular acceptance, the large amount of material
added close to the interaction point will  result in an area
in the calorimeter which will be strongly affected by secondary interactions
and backscattering. This area will have to be excluded from the analysis 
for CC event studies. We took a 
fiducial region of the calorimeter, which amounts to
approximately 10$^\circ$ to 170$^\circ$ in polar angle, considered 
to be  usable for 
the CC analysis. 
To study the effect of
the reduced acceptance, the simulation program was changed accordingly
(but the possible remaining 
influence of backscattering due to the magnets inside
the detectors has not been included). The results show however
that the
quality of the measurement of $A^{W^\pm}$ is not 
significantly affected by the
reduced acceptance, and  is compatible within statistics with the
result of Fig.\ref{fig_results}, using the full acceptance.

\subsection{Influence of the parton density functions}

\begin{figure}[!hp]
\centering
\epsfig{file=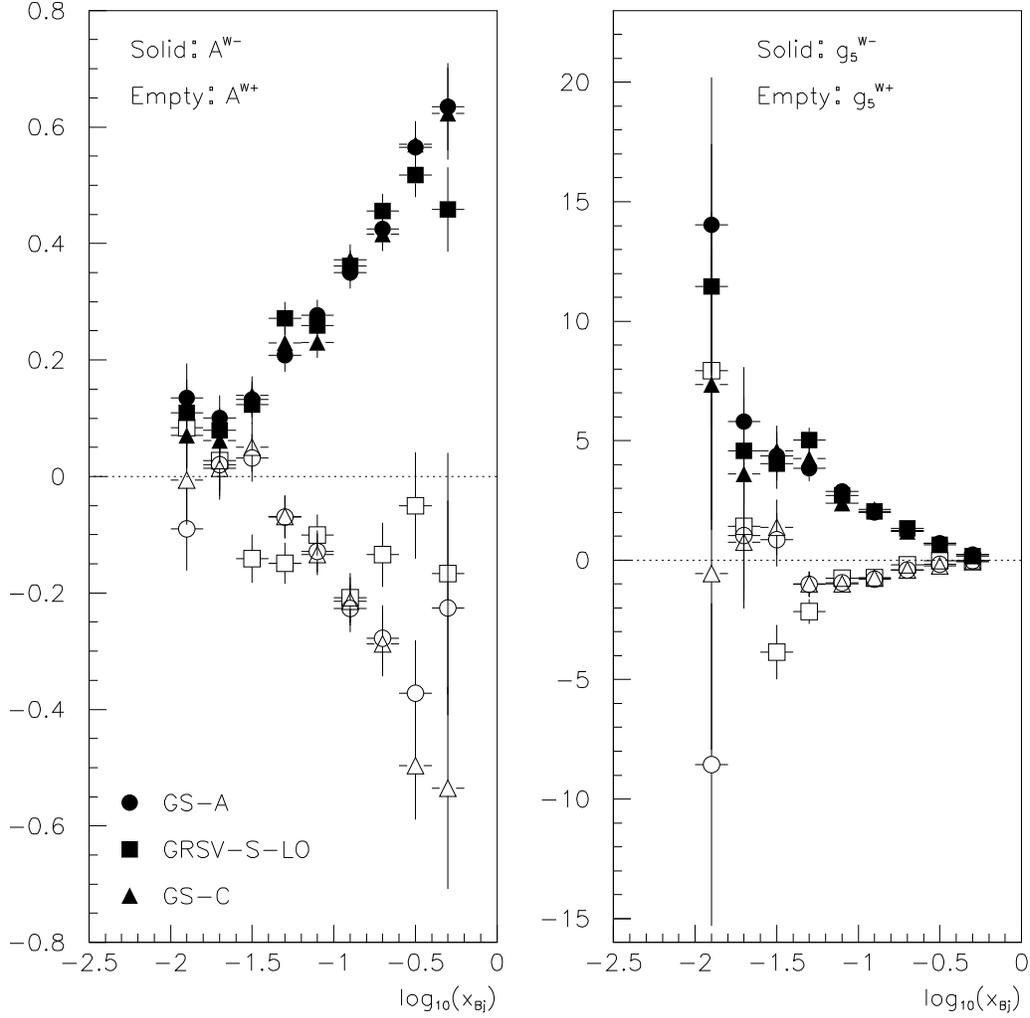, width=.9\textwidth}
\caption{Asymmetries $A^{W^\pm}$ (right) and structure functions
  $g^{W^\pm}_5$ (left) for three different parton densities. In all
  three cases GS--A was used to calculate $F^{W^\pm}_1$, needed to
  extract $g^{W^\pm}_5$ from $A^{W^\pm}$.}
\label{fig_pdf}
\end{figure}

\vspace{1mm}
\noindent
The studies presented so far  have used the parton
densities of GS set A. Here results for GS set C and the parton
densities of reference \cite{grsv} for the standard version in
leading order approximation (labelled GRSV--S--LO) are shown in 
Fig.~\ref{fig_pdf} for $A^{W^\pm}$ and $g^{W^\pm}_5$.
It appears that the predictions are very similar. This is due to 
the limited flavour separation possibilities in extracting these
parton densities (from QCD fits to $g_1$ measurements),
 and in the overall normalization imposed by the $g_1$ measurements.
However, only future measurements will show whether the present assumptions 
are in agreement with reality. 

  To extract
$g^{W^\pm}_5$ (right side of the figure)
the GRV parton densities were used in all three cases
to calculate $F^{W^\pm}_1$.

\section{Extraction of $\Delta u$ and $\Delta d$}

\vspace{1mm}
\noindent
As mentioned in section \ref{sec_intro}, we have in leading order 

\begin{eqnarray*}
g^{W^-}_5 &=& \Delta u+\Delta c - \Delta\bar{d} - \Delta\bar{s}\\
g^{W^+}_5 &=& \Delta d+\Delta s - \Delta\bar{u} - \Delta\bar{c}\\
\end{eqnarray*}

\begin{figure}[!t]
\centering
\epsfig{file=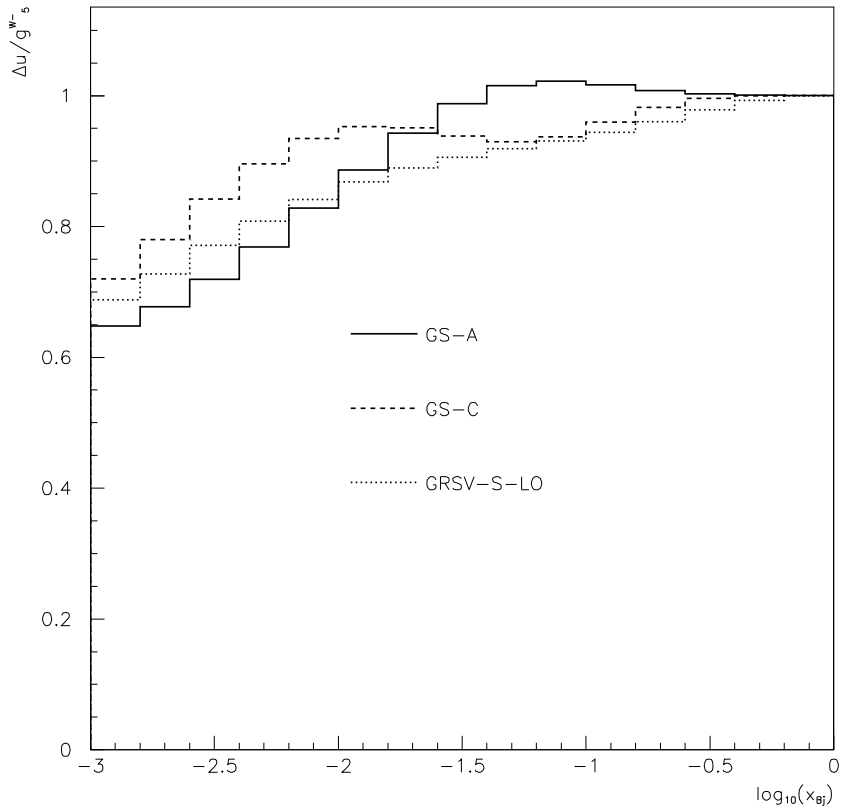,height=.5\textheight}
\caption{Ratio of the $\Delta u$ polarized parton distribution to the
  $g^{W^-}_5$ structure function for three sets of 
polarized parton distributions considered in
  this analysis.}
\label{fig_u_g5}
\end{figure}

In all three parametrizations of the 
polarized parton densities used in this analysis
$\Delta c$ has been neglected. Also the following assumptions are made:
$\Delta\bar{u}=\Delta s$ and $\Delta\bar{d} = \Delta\bar{s}$. 
With these assumptions  the extraction of $g^{W^+}_5$ is equivalent 
 to the
extraction of $\Delta d$. 
Furthermore, $g^{W^-}_5$ is close to $\Delta u$.
This is demonstrated  in Fig.~\ref{fig_u_g5} where the ratio
 of $\Delta u$ to $g^{W^-}_5$ for the three
parametrizations used in this analysis 
 is shown. It is seen that for the kinematic range accessible to
this measurement ($\log_{10}(x)>-2$) all three $\Delta u$
distributions account for more than 85\% of $g^{W^-}_5$.
Hence
the extraction of  $\Delta u$ can be done with a 
precision comparable to the 
 $g^{W^-}_5$ extraction shown in the preceding sections.

\section{Conclusions}

\vspace{1mm}
\noindent

Charged current events, generated with the PEPSI Monte Carlo program, 
have been studied for a polarized HERA. Detector effects were taken into 
account. Charged current events allow a precise  extraction
of  the structure functions 
$g_5^{W^{\pm}}$ in the region of Bjorken-$x$ larger than 0.01. Systematic 
effects
of the calorimeter energy scale, parton distributions
 and detector acceptance were studied, with data samples corresponding to a 
total statistics of 500 pb$^{-1}$. The 
measurable asymmetries remain very prominent
and survive the detector effects.
With simplistic assumptions these structure functions are directly related 
to the polarized quark distributions $\Delta u$ and $\Delta d$. Hence 
the study of asymmetries allows to extract information on these
quantities at a $Q^2$ scale of $O(1000)$ GeV$^2$. 
Although present parton density parametrizations do not 
predict large differences, mainly due to the  limited information 
presently 
available for extracting these densities, this measurement is unique and
will provide vital data on the flavour 
decomposition of the polarized quark densities.
It will be also of interest to check the evolution of these densities 
at low $Q^2$ as measured by SMC and HERMES~\cite{hermes} to the
kinematic region of HERA. 

\section{Acknowledgements}

\vspace{1mm}
\noindent

 We thank 
J. Kalinowski
for helpful discussions.
J.G.C.    acknowledges the support by Bundesministerium f\"ur 
Bildung, Wissentschaft, Forschung
und Technologie, FRG contract 6DO571
   and  the Schoolarship from the Graduiertenkolleg DFG.

\noindent
``These Proceedings'' refers to the Proceedings of the Workshop on Physics
at HERA with Polarized Protons and Electrons.

\end{document}